\documentclass[a4paper,fleqn]{cas-sc}
\usepackage{diagbox}
\usepackage[authoryear,longnamesfirst]{natbib}
\usepackage{wrapfig}
\usepackage[ruled,vlined]{algorithm2e}
\usepackage{algpseudocode} 
\usepackage{graphicx}
\graphicspath{ {Figures/} }
\usepackage{caption}
\usepackage{subcaption}
\usepackage{hyperref}

\def\tsc#1{\csdef{#1}{\textsc{\lowercase{#1}}\xspace}}
\tsc{WGM}
\tsc{QE}
\tsc{EP}
\tsc{PMS}
\tsc{BEC}
\tsc{DE}

\begin{document}
\let\WriteBookmarks\relax
\def\floatpagepagefraction{1}
\def\textpagefraction{.001}
\shorttitle{}

\title [mode = title]{Exploring Structural Dynamics in Retracted and Non-Retracted Author's Collaboration Networks: A Quantitative Analysis}                      

\author[1,2]{Kiran Sharma}\corref{cor1}
\ead{kiran.sharma@bmu.edu.in}
\author[1]{Aanchal Sharma}
\author[1]{Jazlyn Jose}
\author[1]{Vansh Saini}
\author[1]{Raghavraj Sobti}
\author[1,2]{Ziya Uddin}
%\credit{Conceived and designed the analysis; Collected the data;  Performed the analysis; Writing - original draft}

\address{School of Engineering \& Technology, BML Munjal University, Gurugram, Haryana-122413, India }
\address{Center for Advanced Data and Computational Science, BML Munjal University, Gurugram, Haryana-122413, India }

\cortext[cor1]{Corresponding author}
%=========================================================
\begin{abstract}
Retractions compromise the reliability of the scientific literature and affect the foundation of further research. Understanding the structure of collaboration networks in retracted papers helps identify risk factors, such as recurring co-authors or institutions. To compare the network structures of retracted and non-retracted networks, we selected 30 authors with a sufficiently large number of retractions. The number of retracted papers for each author was obtained from the Retraction Watch database, while non-retracted papers were sourced from the Scopus database. Using this data, collaboration networks for retracted and non-retracted authors were constructed. To analyze the dynamics between these networks, various network properties were measured. As a result, retracted collaboration networks show more hierarchical and centralized structures, with strong correlations between degree and centrality measures, whereas non-retracted collaboration network metrics emphasize distributed collaboration with strong clustering and connectivity, indicating more balanced network structures compared to retracted metrics. The comparison between retracted and non-retracted collaboration networks highlights structural and functional overlaps. This indicates consistency in clustering and centrality across retracted and non-retracted networks, but highlights differences in weighted connections and assortative mixing. For statistical validation among the network metrics, a $t$-test is used to determine whether there is a statistically significant difference between the means of two groups. Metrics such as Degree Centrality, Average Weighted Degree, and Closeness Centrality reveal significant structural differences between the networks. These results highlight which aspects of the network topology differ significantly, shedding light on how structural and dynamic properties can vary between the two contexts. Finally, Cohen's $d$ test is used after a $t$-test to quantify the effect size, providing a standardized measure of the magnitude of the difference between two groups beyond statistical significance. Hence, understanding how retraction-prone collaborations form can inform policies to improve research practices.
\end{abstract}

\maketitle
%=========================================================
\section{Introduction}

Retractions pose serious challenges to scientific integrity, reputation, and the broader scientific enterprise. They undermine the credibility of research by revealing some type of error, data manipulation, or ethical violations that create distrust in the science process and the presented findings~\citep{Fanelli2013}. Retractions also destroy authors' and institutions' reputations, which often result in professional sanctions, a decrease in funding opportunities, and long-term damage to their credibility within the academic community~\citep{Lu2023, sharma2024ripple}. In addition, if retractions are not properly flagged, they can mislead future research by allowing flawed or falsified findings to remain in the literature, amplifying the damage to the scientific record. This can be problematic for trust, helps with the spread of misinformation, thus hindering scientific progress. As such, the varied public perception of science remains influenced, though in truth, it is life-critical in fields such as medicine and public health. The resolution of these challenges calls for proactive monitoring and analysis; network science brings rich tools to help investigate the patterns and consequences of retractions.

Network science has emerged as a transformational tool in the analytical investigation of growth and structure in social systems~\citep{sharma2021growth}. It has allowed deep insight, with the applications of mathematical, computational, and visualization techniques, into how individuals and organizations connect with each other, build their relationships out, and influence one another~\citep{Barabasi2016}.Growth patterns, such as preferential attachment, explain how influential researchers or institutions attract more collaborations, creating hubs that shape the structure of science~\citep{khurana2024growth}.T
This method helps uncover key contributors, interdisciplinary links, and emerging patterns in collaborative research, providing critical insights into the growth and efficiency of scientific ecosystems.

It was seen by \cite{Barabási2001Evolution} that collaborations form in the research landscape change over time and analyzed that these networks show scale free properties where there are a few authors that serve as hubs connecting with many authors while others maintain fewer links compared to them.  The preferential attachment mechanism, which describes the tendency of new authors to connect with already well-connected collaborators, underpins this network structure. As a result, networks show increasing average connectivity and decreasing average separation, aligning with the small-world phenomenon, characterized by high clustering coefficients and short average path lengths. It was seen that as time has progressed the co-authorship networks have become more connected and denser that reflects broader trends in the scientific collaboration,  These findings These findings highlight the importance of understanding collaboration dynamics to analyze the impacts of retractions and shifts in co-authorship networks.

Further, the study by \cite{Yan2009Applying} focused on evaluating the centrality measures to assess the academic impact of authors in the LIS field. A co-authorship network was constructed and properties like mean distance, clustering coefficient, and the largest component were analyzed. Centrality measures like degree, closeness, betweenness, and PageRank were applied to rank authors and correlate their positions with citation counts. This study’s findings indicated that there was a significant correlation between centrality measures and citation counts with betweenness centrality showing high relationship. This study showed that centrality measures can provide valuable insights into the article impact.This study suggested that combining centrality measures can provide for a holistic academic evaluation.

%=========================================================
\section{Literature review}
Co-authorship networks uncover patterns of scientific collaboration, highlighting variations across disciplines and changes over time~\citep{Newman2004Coauthorship}. However, scientific misconduct can significantly impact collaboration networks, affecting both the individuals directly involved and their prior collaborators. 

\cite{Hussinger2018Guilt} shows that scientific misconduct imposes significant costs on unsuspecting past collaborators, whose citations fall by 8-9\% relative to a control group. This is most likely due to stigmatization by association, which incurs considerable indirect costs of misconduct, including generalized distrust in research. Such backlash could raise the stakes for whistleblowers when exposing misconduct in order to protect their name and career.
~\cite{sharma2024ripple} highlighted in their study that retractions in scientific research can have significant impacts on the collaboration networks of the authors involved. They examined whether retractions due to scientific misconduct can lead to stigmatization and isolation for the authors involved, affecting their collaboration networks. Co-authors also face significant career impacts, particularly those listed as the first or last authors. Although retracted authors who continue to publish may maintain and establish new collaborations, these are often with less senior and less impactful co-authors. 

Further, the study by \citep{sharma2021team} investigates the patterns of retracted scientific publications over nearly four decades, focusing on the association between the collaboration of the authors, the retracted articles, the retracted citations, the impact factor of the journal, and the research areas. The analysis reveals that scientific misconduct has a ripple effect, leading to further retractions and impacting new research. In addition, \citep{Jin2019The} showed that the retraction of scientific teams leaves the previous work of the eminent co-authors undamaged but significantly decreases the citations of the less eminent co-authors. Furthermore, according to the findings presented in the work by~\citep{Azoulay2015The}, for the retracted authors, the scientific retraction implies a 10\% drop in citation rates; prominent scientists are penalized much more than their less distinguished colleagues. 

Scientific misconduct, including plagiarism, data fabrication, and falsification, is a significant issue in the academic community around the world. This misconduct not only undermines the integrity of scientific research, but also erodes public trust in science. At least 10\% of scientists observed scientific misconduct in their career~\citep{Gross2016Scientific}. \cite{Steen2013Why} concluded that both changes in the behavior of authors and changes in the policies and efficiency of institutions are responsible for the increased rate of retractions. They note that the rise in retractions is not solely due to an increase in misconduct but also reflects a lowering of barriers to the publication and subsequent identification and retraction of flawed articles. During the last two decades, numerous studies have investigated the causes of retractions and evaluated the quality of journals involved in cases of scientific misconduct. \cite{sharma2024over} showed that scientific misconduct remains a major issue in India, with plagiarism, data fabrication, and falsification being the leading causes of retractions. The quality of the journals and the nature of the collaboration play a crucial role in the appearance of retractions. Strengthening institutional policies and fostering a supportive academic culture are essential steps toward mitigating scientific misconduct and ensuring the integrity of scientific research. 

Scientific misconduct is more likely to occur in countries that lack research integrity policies, offer cash rewards for publication, within cultures that have limited mutual criticism, and at earlier stages of the career, and are not related to gender or pressure to publish~\citep{Fanelli2015Misconduct}.
\cite{Candal-Pedreira2023Research} discussed the different measures that academia and scientific journals have focused on to overcome the challenge of research misconduct and puts into sharp light the fact that it is a multi-stakeholder effort that alone shall result in preventing the negative consequences due to research misconduct and public distrust in science.

%=========================================================
\subsection{Research gap}
This study highlights a gap in the comprehensive understanding of the structural differences and dynamics between collaboration networks involving retracted and non-retracted works. Despite previous analyses, the specific role of network properties such as centrality, clustering, and transitivity in influencing retraction-prone behaviors remains unexplored.
%=========================================================
\subsection{Research objectives}
The research objectives are two-fold.
\begin{enumerate}
    \item Examine the structural differences in authors' collaboration networks between retracted and non-retracted publications.
    \item Evaluate the statistical significance and effect size of these differences using advanced metrics such as $t$-test and Cohen's $d$ test.
\end{enumerate}
%=========================================================
\section{Data description}

The datasets analyzed in this study comprise two main categories: retracted and non-retracted co-authorship records. The data were sourced from CrossRef~\href{https://www.crossref.org/blog/news-crossref-and-retraction-watch/}, Retraction Watch~\href{https://retractionwatch.com/the-retraction-watch-leaderboard/}, and Scopus~\href{https://www.scopus.com/search/}. These datasets were curated to facilitate a comparative analysis of network structures between retracted and non-retracted publications.
%~~~~~~~~~~~~~~~~~~~~~~~~~~~~~~~~~~~~~~~~~~~~~~~~~~~~~~~~~~~~~~~~~~~~~~~
\subsection{Data collection}

Retracted papers were sourced from the CrossRef Retraction Database, which contained 56,330 records. For the top 30 authors listed on the Retraction Watch Leaderboard, ranked by the number of their retracted papers, retracted publications were filtered from the CrossRef database. Details such as publication data, co-authors, affiliations, publication types, and citation metrics were extracted for each author. This data was downloaded on September 23, 2024. In contrast, the non-retracted dataset includes all other publications by the same authors that have not been retracted, as identified through the Scopus database.

\subsection{Data pre-processing}

Since our research focuses exclusively on the top 30 authors, we filtered the Crossref dataset to create a subset containing papers published only by these 30 authors. Upon cross-referencing the Retraction Watch leaderboard with the CrossRef dataset, we found that six authors were present only in the retracted dataset and not in the non-retracted records (Scopus). As a result, these six authors were excluded from the non-retracted dataset, leaving 24 authors for further analysis. 

The excluded authors, along with their number of retractions in Scopus, are A. Salar Elahi (76), Chen-Yuan (Peter) Chen (43), Hua Zhong (41), James Hunton (36), Antonio Orlandi (34), and Dimitris Liakopoulos (33). Furthermore, there were complications with name variations that created ambiguity in matching authors across datasets. As a result, Amelec Viloria was excluded due to multiple name variations. Additionally, Jun Ren and Bharat Agarwal were removed because their publications frequently involved an unusually high number of co-authors—which posed significant computational challenges for the analysis. Lastly, Hua Zhong was excluded due to persistent network issues that hindered data retrieval and processing. After these exclusions, the dataset was reduced to 20 authors from the original top 30 leaderboard.

To maintain a consistent dataset of 30 authors, 10 additional authors were then selected from the top 100 list of the CrossRef Retraction Database. These authors, who also had available Scopus IDs for their non-retracted publications, were added to the dataset, ensuring both retracted and non-retracted publications were represented. The additional authors included Annarosa Leri (49 retractions), Ekta Roy (20 retractions), Gunter G. Hempelmann (97 retractions), Hans J. Eysenck (93 retractions), Hironobu Ueshima (125 retractions), Hiroyoshi Tanaka (118 retractions), Li Zhang (75 retractions), Pattium Chiranjeevi (20 retractions), Saber Khelaifia (49 retractions), and Stefan W. Suttner (47 retractions). With these additions, the final dataset comprised 30 authors, providing a well-rounded base for analysis. This selection ensured a consistent focus on authors with both retracted and non-retracted works, facilitating a direct comparison. 

Additionally, Retraction Watch does not provide Scopus IDs for authors. To obtain the Scopus IDs needed to retrieve the authors' non-retracted profiles from Scopus, we first identified retracted papers along with their corresponding PubMed IDs from Retraction Watch. Using these PubMed IDs, we located the authors in the Scopus database. By querying Scopus with the identified Scopus author IDs, we downloaded the individual profiles for the 30 authors. Table \ref{table:retracted_authors} displays the list of 30 authors along with their retracted papers count and non-retracted papers count.
%~~~~~~~~~~~~~~~~~~~~~~~~~~~~~~~~~~~~~~~~~~~~~~~~~~~~~~~~~~~~~~~~~~~~~~~

%~~~~~~~~~~~~~~~~~~~~~~~~~~~~~~~~~~~~~~~~~~~~~~~~~~~~~~~~~~~~~~~~~~~~~~~
\begin{table}[!ht]
\centering
\caption{List of 30 authors as per the number of retracted and their corresponding non-retracted papers.}
 \resizebox{\textwidth}{!}{ 
\begin{tabular}{|c|l|cc|c|l|cc|}
\hline
\multirow{2}{*}{\textbf{S. No.}} & \multirow{2}{*}{\textbf{Author Name}} & \multicolumn{2}{c|}{\textbf{Number of Papers}}                   & \multirow{2}{*}{\textbf{S. No.}} & \multirow{2}{*}{\textbf{Author Name}} & \multicolumn{2}{c|}{\textbf{Number of Papers}}                   \\ \cline{3-4} \cline{7-8} 
                                 &                                       & \multicolumn{1}{c|}{\textbf{Retracted}} & \textbf{Non-retracted} &                                  &                                       & \multicolumn{1}{c|}{\textbf{Retracted}} & \textbf{Non-retracted} \\ \hline
1                                & Ali Nazari                            & \multicolumn{1}{c|}{110}                & 133                    & 16                               & Jun Iwamoto                           & \multicolumn{1}{c|}{110}                & 158                    \\ \hline
2                                & Annarosa Leri                         & \multicolumn{1}{c|}{49}                 & 142                    & 17                               & Li Zhang                              & \multicolumn{1}{c|}{75}                 & 192                    \\ \hline
3                                & Ashok Pandey                          & \multicolumn{1}{c|}{44}                 & 679                    & 18                               & Naoki Mori                            & \multicolumn{1}{c|}{31}                 & 226                    \\ \hline
4                                & Diederik A. Stapel                    & \multicolumn{1}{c|}{58}                 & 81                     & 19                               & Pattium Chiranjeevi                   & \multicolumn{1}{c|}{20}                 & 44                     \\ \hline
5                                & Dong Mei Wu                           & \multicolumn{1}{c|}{44}                 & 81                     & 20                               & Prashant K Sharma                     & \multicolumn{1}{c|}{31}                 & 113                    \\ \hline
6                                & Ekta Roy                              & \multicolumn{1}{c|}{20}                 & 26                     & 21                               & Saber Khelaifia                       & \multicolumn{1}{c|}{49}                 & 120                    \\ \hline
7                                & Fazlul H Sarkar                       & \multicolumn{1}{c|}{53}                 & 531                    & 22                               & Shahaboddin Shamshirband              & \multicolumn{1}{c|}{51}                 & 572                    \\ \hline
8                                & Gunter G Hempelmann                   & \multicolumn{1}{c|}{97}                 & 691                    & 23                               & Shigeaki Kato                         & \multicolumn{1}{c|}{44}                 & 502                    \\ \hline
9                                & Hans J Eysenck                        & \multicolumn{1}{c|}{93}                 & 465                    & 24                               & Soon-Gi Shin                          & \multicolumn{1}{c|}{30}                 & 44                     \\ \hline
10                               & Hironobu Ueshima                      & \multicolumn{1}{c|}{125}                & 27                     & 25                               & Stefan W. Suttner                     & \multicolumn{1}{c|}{47}                 & 84                     \\ \hline
11                               & Hiroyoshi Tanaka                      & \multicolumn{1}{c|}{118}                & 90                     & 26                               & Victor Grech                          & \multicolumn{1}{c|}{93}                 & 298                    \\ \hline
12                               & Hyung-In Moon                         & \multicolumn{1}{c|}{31}                 & 118                    & 27                               & Yogeshwer Shukla                      & \multicolumn{1}{c|}{17}                 & 160                    \\ \hline
13                               & Jan Hendrik Schon                     & \multicolumn{1}{c|}{21}                 & 112                    & 28                               & Yoshihiro Sato                        & \multicolumn{1}{c|}{159}                & 130                    \\ \hline
14                               & Joachim Boldt                         & \multicolumn{1}{c|}{233}                & 408                    & 29                               & Yoshitaka Fujii                       & \multicolumn{1}{c|}{220}                & 194                    \\ \hline
15                               & Jose Luis Calvo-Guirado               & \multicolumn{1}{c|}{33}                 & 164                    & 30                               & Yuhji Saitoh                          & \multicolumn{1}{c|}{66}                 & 60                     \\ \hline
\end{tabular}
}
\label{table:retracted_authors}
\end{table}
%~~~~~~~~~~~~~~~~~~~~~~~~~~~~~~~~~~~~~~~~~~~~~~~~~~~~~~~~~~~~~~~~~~~~~~~
%=========================================================
\section{Methodology}

\subsection{Path length} 
This is the measure of distance between nodes in the network expressed in terms of the number of edges traversed while connecting two nodes. It is the number of intermediate co-authors that would connect two authors in the co-authorship network through their collaborative relationships.

Equation~\ref{eq:PL} is the mathematical formulation to calculate average path length.
\begin{equation}
L = \frac{1}{n(n-1)} \sum_{i \neq j} d(i, j)
\label{eq:PL}
\end{equation}
where \(L\) is the average path length, \(d(i,j)\) is the shortest path between nodes \(i\) and \(j\), and \(n\) is the number of nodes in the network. Short path length means that authors are highly interconnected, thereby increasing the speed of knowledge exchange. Long path length implies a greater distance that separates authors into more distantly related groupings.
%~~~~~~~~~~~~~~~~~~~~~~~~~~~~~~~~~~~~~~~~~~~~~~~~~~~~~~~~~~~~~~~~~~~~~~~
\subsection{Weighted edges} 
In a network, weighted edges are those where the connections (or edges) between nodes have an associated weight of the relationship. For two nodes \(i\) and \(j\), if they are connected by an edge with weight \(w_{ij}\), the adjacency matrix \(A\) can be replaced by a weight matrix \(W\).

\begin{equation}
W = [w_{ij}]
\label{eq:WE}
\end{equation}
where \( w_{ij} \) is the weight of the edge between node \(i\) and node \(j\).
%~~~~~~~~~~~~~~~~~~~~~~~~~~~~~~~~~~~~~~~~~~~~~~~~~~~~~~~~~~~~~~~~~~~~~~~
\subsection{Clustering coefficient} 
The clustering coefficient indicates the degree to which a node's neighbors are interconnected, revealing the probability that two neighbors of a node are also connected to each other. For a node \(v\), the local clustering coefficient can be calculated as (Eq\ref{eq:CC})

\begin{equation}
C(v) = \frac{2e_v}{k_v(k_v - 1)}
\label{eq:CC}
\end{equation}
where \(e_v\) is the number of edges between the neighbors of node \(v\) and \(k_v\) is the degree of node \(v\) (i.e., the number of neighbors).

The global clustering coefficient is the average of the local clustering coefficients across all nodes, and can be calculated as (Eq\ref{eq:GCC})

\begin{equation}
C_{global} = \frac{1}{n} \sum_{v} C(v)
\label{eq:GCC}
\end{equation}
where \(n\) is the total number of nodes in the network.
%~~~~~~~~~~~~~~~~~~~~~~~~~~~~~~~~~~~~~~~~~~~~~~~~~~~~~~~~~~~~~~~~~~~~~~~
\subsection{Assortativity}

Assortativity refers to the tendency of nodes in a network to connect with other nodes that are similar (or dissimilar) in some attribute. In network science, assortativity is commonly quantified using the assortativity coefficient. It is calculated based on the attributes of connected nodes, such as degree (degree assortativity). The coefficient can range from -1 (perfect disassortativity, where nodes prefer to connect with very different nodes) to 1 (perfect assortativity, where nodes prefer to connect with similar nodes). A value near 0 indicates no particular preference. The degree assortativity coefficient, $r$, is given by:

\begin{equation}
r = \frac{\sum_{i} (j_i - \bar{j})(k_i - \bar{k})}{\sqrt{\sum_{i} (j_i - \bar{j})^2 \sum_{i} (k_i - \bar{k})^2}}
\end{equation}

where \( j_i \) and \( k_i \) are the degrees of the nodes at the ends of the \( i \)-th edge.
\( \bar{j} \) and \( \bar{k} \) are the mean degrees of the nodes across all edges.

%~~~~~~~~~~~~~~~~~~~~~~~~~~~~~~~~~~~~~~~~~~~~~~~~~~~~~~~~~~~~~~~~~~~~~~~
\subsection{Transitivity}

Transitivity is a global network measure that captures the overall tendency of the network to form triangles. It does not focus on individual nodes but considers the network as a whole. It quantify the level of clustering or cohesiveness in a graph. A network with high transitivity tends to have tightly-knit communities. The transitivity coefficient, often referred to as the global clustering coefficient, is given by:
\begin{equation}
    T = \frac{3 \times \text{number of closed triangles}}{\text{number of connected triples of nodes}}
\end{equation}
%~~~~~~~~~~~~~~~~~~~~~~~~~~~~~~~~~~~~~~~~~~~~~~~~~~~~~~~~~~~~~~~~~~~~~~~
\subsection{Centrality measures}
Centrality measures are key metrics in network analysis used to identify the most important or influential nodes within a network based on their connectivity and position relative to other nodes. 
%~~~~~~~~~~~~~~~~~~~~~~~~~~~~~~~~~~~~~~~~~~~~~~~~~~~~~~~~~~~~~~~~~~~~~~~
\subsubsection{Degree centrality}
Degree centrality is the simplest centrality measure and measures the number of nodes a certain node is connected to. This helps us infer the most connected node in the network other than the main author and can be calculated as (Eq~\ref{eq:DC})

\begin{equation}
C_D(v) = \deg(v)
\label{eq:DC}
\end{equation}
where \( \deg(v) \) is the degree (number of edges) of node \(v\).
%~~~~~~~~~~~~~~~~~~~~~~~~~~~~~~~~~~~~~~~~~~~~~~~~~~~~~~~~~~~~~~~~~~~~~~~
\subsubsection{Betweenness centrality}
This quantifies the number of times a node acted as a bridge in the shortest path between two nodes. It is crucial for understanding which nodes are responsible for controlling the information flow between nodes and can be calculated as (Eq~\ref{eq:BC})

\begin{equation}
C_B(v) = \sum_{s \neq v \neq t} \frac{\sigma_{st}(v)}{\sigma_{st}}
\label{eq:BC}
\end{equation}
where \( \sigma_{st} \) is the total number of shortest paths from node \(s\) to node \(t\) and \( \sigma_{st}(v) \) is the number of shortest paths from \(s\) to \(t\) that pass through node \(v\).
%~~~~~~~~~~~~~~~~~~~~~~~~~~~~~~~~~~~~~~~~~~~~~~~~~~~~~~~~~~~~~~~~~~~~~~~
\subsubsection{Closeness centrality}
Closeness centrality shows how close a node is to all other nodes on the network, based on shortest path. It is reciprocal of the sum of shortest path to all the nodes and can be calculated as (Eq~\ref{eq:CoC})

\begin{equation}
C_C(v) = \frac{1}{\sum_{t} d(v,t)}
\label{eq:CoC}
\end{equation}
where \(d(v,t)\) is the shortest distance between node \(v\) and node \(t\).
%~~~~~~~~~~~~~~~~~~~~~~~~~~~~~~~~~~~~~~~~~~~~~~~~~~~~~~~~~~~~~~~~~~~~~~~
\subsubsection{Eigenvector centrality}
The influence of a node in a network is measured by eigenvector centrality, which assigns higher scores to nodes that are connected to other high-scoring nodes. It can be calculated as (Eq~\ref{eq:EC})

\begin{equation}
    C_E(v) = \frac{1}{\lambda} \sum_{u \in N(v)} A_{vu} C_E(u)
    \label{eq:EC}
\end{equation}
where \( \lambda \) is a constant (eigenvalue), \( A_{vu} \) is the adjacency matrix entry between nodes \(v\) and \(u\), \( C_E(u) \) is the eigenvector centrality of node \(u\), and \( N(v) \) is the set of neighbors of node \(v\).
%~~~~~~~~~~~~~~~~~~~~~~~~~~~~~~~~~~~~~~~~~~~~~~~~~~~~~~~~~~~~~~~~~~~~~~~
%%%##################################################################

%%%##################################################################
\section{Results and discussion}

\subsection{Author's collaboration network}
\begin{figure}[!ht]
    \centering
\includegraphics[width=0.95\linewidth]{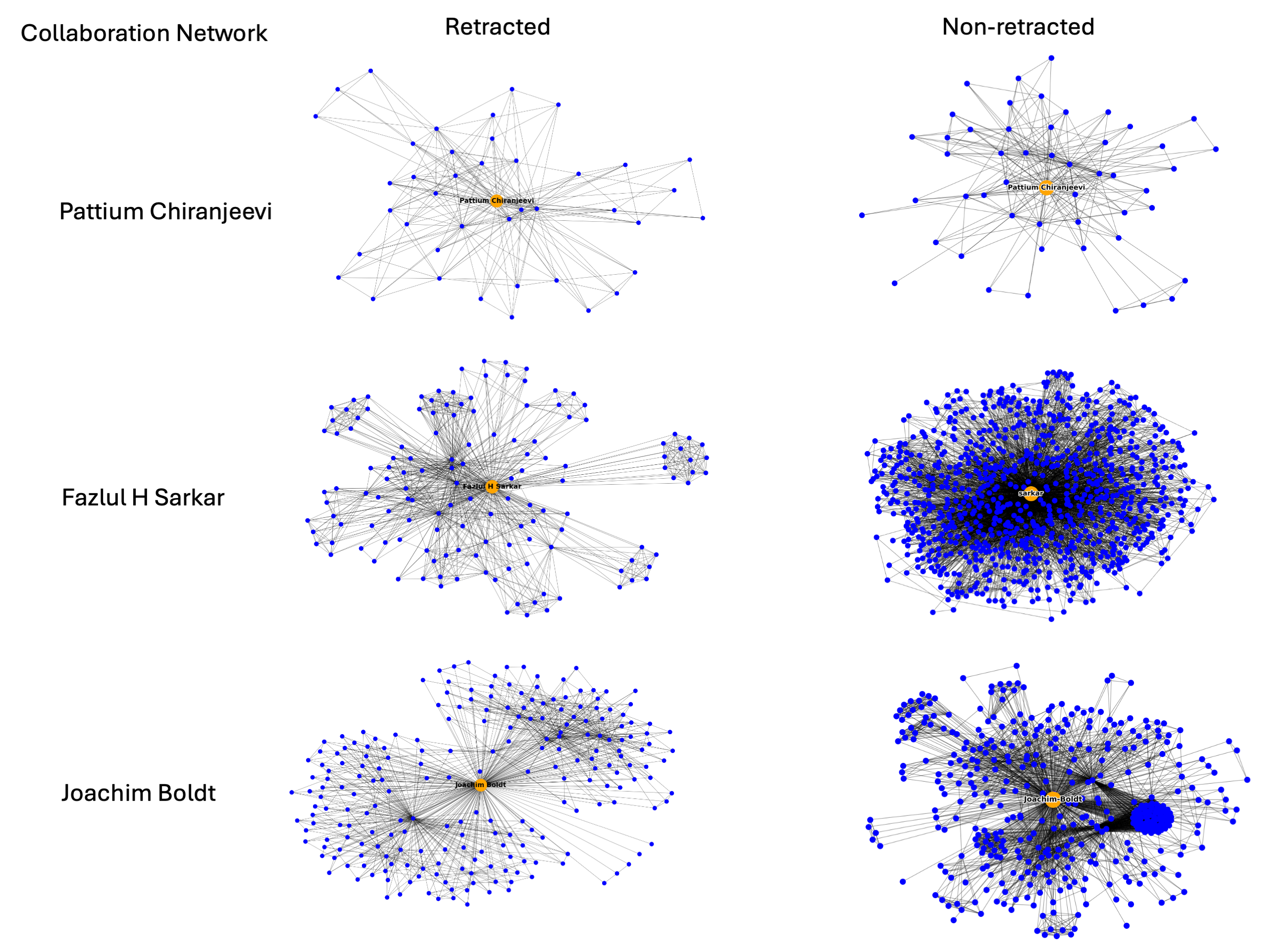} 
\caption{Author's collaboration network of retracted and non-retracted publications. Demonstration of three authors.}
\label{fig:network}
\end{figure}

Authors' collaboration networks represent relationships and interactions between researchers, typically visualized as graphs where nodes correspond to authors and edges represent co-authored publications~\citep{newman2000structure}. These networks help to analyze how researchers collaborate within and across disciplines~\citep{Newman2004Coauthorship}.
Figure~\ref{fig:network} highlights the collaboration network of three selected authors from the list of 30 authors (see Table~\ref{table:retracted_authors}). These authorship collaboration networks for both retraction and non-retraction shows a different networks structure. 

%~~~~~~~~~~~~~~~~~~~~~~~~~~~~~~~~~~~~~~~~~~~~~~~~~~~~~~~~~~~~~~~~~~~~~~~
%~~~~~~~~~~~~~~~~~~~~~~~~~~~~~~~~~~~~~~~~~~~~~~~~~~~~~~~~~~~~~~~~~~~~~~~

%=========================================================
\subsection{Network metrics}
Network properties are computed to gain a deeper understanding of the structure, dynamics, and behavior of the system, enabling the identification of patterns, influential entities, and structural differences, such as collaboration networks~\citep{furukawa2011mobility}. Node-level metrics, such as degree centrality and eigenvector centrality, are useful for identifying influential nodes, while global properties like average path length and network density characterize the network's overall connectivity and efficiency. These metrics are crucial for identifying key entities, analyzing community structures, and comparing dynamics between different networks, such as those of retracted and non-retracted collaborations. For example, examining clustering coefficients and modularity can reveal how nodes group into communities or how centralized the network is, offering insights into collaboration patterns. 

In addition, network properties enable the evaluation of robustness and resilience, helping researchers understand how a network reacts to node failures or targeted removals. Such analyses also allow comparisons between real-world networks and theoretical models, providing a basis for improving systems or identifying structural vulnerabilities \citep{newman2003structure, barabasi2016network}. Linking these properties to real-world outcomes reveals how network structure impacts collaboration efficiency, resilience, and retractions in academic networks. For all 30 authors, the properties of the collaboration network for retracted publications are specified in Table~\ref{table:RetProperty} and for non-retracted papers are shown in Table~\ref{table:nonretProperty}.

%=========================================================
\subsection{Network metrics comparison}

Examining the correlation between network properties enables researchers to understand how structural patterns influence real-world phenomena, such as collaboration efficiency, information dissemination, or the likelihood of retractions in academic networks. For instance, a positive correlation between degree centrality and betweenness centrality may indicate that highly connected nodes also act as critical bridges within the network, facilitating communication between different regions. Similarly, the relationship between clustering coefficients and average path lengths can shed light on how local connectivity impacts global efficiency, especially in social or collaboration networks. By analyzing these correlations, researchers can uncover dependencies or independence among network properties. This understanding provides insights into the structural organization of the network and the roles played by individual nodes~\citep{newman2003structure, barabasi2016network}.

To understand the similarity between the network structures and properties of retracted and non-retracted authors' collaboration networks, the correlation between the network properties of both types has been calculated. Figure~\ref{fig:CorrPlot} shows the correlation plot for network metrics (a) retracted vs. retracted, (b) non-retracted vs. non-retracted, and (c) retracted vs. non-retracted. 

%%~~~~~~~~~~~~~~~~~~~~~~~~~~~~~~~~~~~~~~~~~~~~~~~
\subsubsection{Correlation among retracted metrics}
Figure \ref{fig:CorrPlot} (left) shows that closeness centrality is highly correlated with degree centrality, indicating that nodes with many direct connections (high DC) also tend to have high closeness centrality, emphasizing their role in efficiently connecting to other nodes. Transitivity (global clustering) and clustering coefficient (local clustering) are strongly linked, suggesting that nodes in highly clustered neighborhoods contribute to overall network clustering. Eigenvector centrality aligns with degree centrality, reflecting that well-connected nodes are also central in terms of influence. On the other hand, a perfect negative correlation between average path length and degree centrality implies that networks with high connectivity (high DC) tend to have shorter average paths, indicating efficient communication. In addition, Nodes with high betweenness centrality (acting as bridges) tend to have lower clustering coefficients, as such nodes often connect disparate parts of the network. Overall, retracted metrics show dense relationships between centrality measures and clustering, revealing hierarchical or hub-like structures in retracted networks.
%%~~~~~~~~~~~~~~~~~~~~~~~~~~~~~~~~~~~~~~~~~~~~~~~
\subsubsection{Correlation among non-retracted metrics}
Similar to retracted metrics, closeness centrality strongly correlates with degree centrality, highlighting the importance of highly connected nodes in reducing path lengths. Betweenness centrality has a stronger positive correlation with degree centrality compared to retracted networks, suggesting a more distributed influence among well-connected nodes. Transitivity correlates strongly with assortativity, reflecting cohesive subnetworks (tightly-knit groups or clusters of nodes) in non-retracted collaborations. However, the same perfect negative correlation between average path length and degree centrality indicates that connectivity remains inversely related to average path length. In addition, eigenvector centrality shows a moderate negative correlation with weighted degree, suggesting that nodes with high weights (intensity of connections) might not always align with high influence in global terms. Overall, non-retracted metrics emphasize distributed collaboration with strong clustering and connectivity, indicating more balanced network structures compared to retracted metrics.
%%~~~~~~~~~~~~~~~~~~~~~~~~~~~~~~~~~~~~~~~~~~~~~~~
\subsubsection{Correlation among retracted and non-retracted metrics}
Clustering coefficient in retracted and non-retracted networks shows a strong positive correlation, indicating that clustering tendencies are preserved across both types of networks. Also, closeness centrality remains consistently correlated between retracted and non-retracted metrics, emphasizing that central nodes in one network type tend to retain their influence in the other. Betweenness centrality is moderately correlated, suggesting that bridging nodes in retracted networks often remain influential in non-retracted networks. Weighted degree shows a weak negative correlation, suggesting that retracted and non-retracted networks may differ in terms of connection intensity. Further, assortativity shows a moderate negative correlation, indicating structural differences in how similar nodes connect across the two network types. Overall, this highlights the structural and functional overlaps between retracted and non-retracted networks while also identifying key differences, such as the reduced role of weighted connections in retracted networks.

%~~~~~~~~~~ correlation chart~~~~~~~~~~~~~~~~~~~~~~~~~~~~~~~~~~~~~~~~~~~~~~~~~~~~~~~~~~~~~
\begin{figure}[!ht]
    \centering
\includegraphics[width=\linewidth]{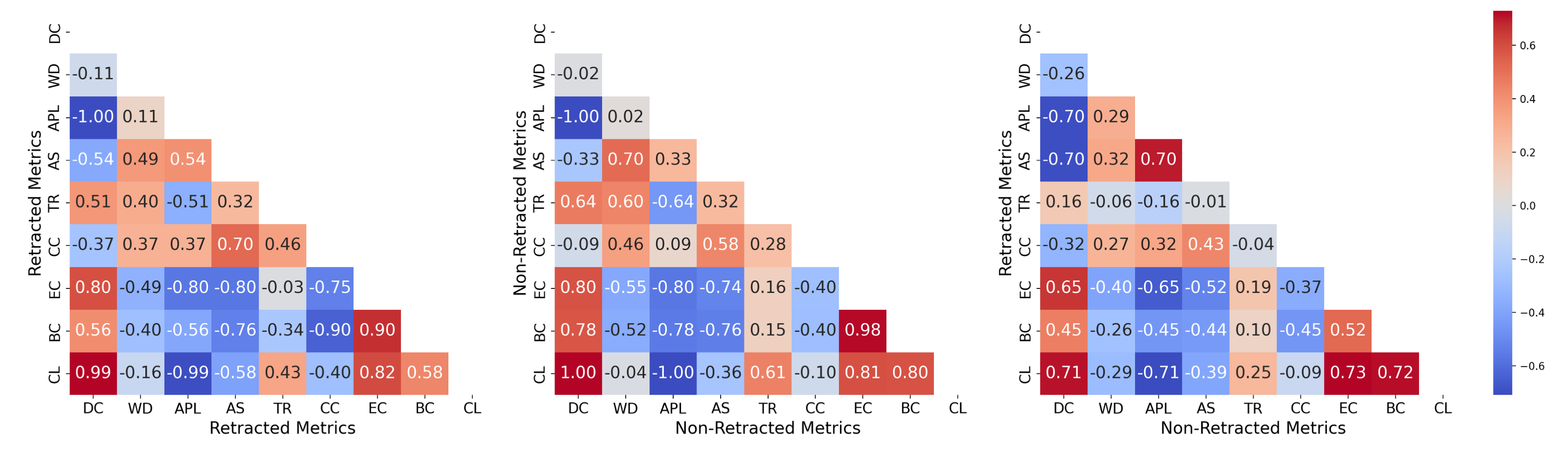} 
\caption{Correlation plot of network metrics. (\textit{left}) Retracted only (\textit{middle}) Non-retracted only, and (\textit{right}) Retracted vs Non-retracted.}
\label{fig:CorrPlot}
\end{figure}
%~~~~~~~~~~~~~~~~~~~~~~~~~~~~~~~~~~~~~~~~~~~~~~~~~~~~~~~~~~~~~~~~~~~~~~~

Figure~\ref{fig:CorrPlot} presents three heatmaps representing correlation matrices for different sets of metrics related to network or graph properties. Each heatmap uses a color scale, where red indicates strong positive correlations (close to 1), blue indicates strong negative correlations (close to -1), and white/gray represents weak or no correlation (close to 0). The rows and columns are labeled with metrics such as DC (Degree Centrality), WD (Weighted Degree), APL (Average Path Length), AS (Assortativity), TR (Transitivity), CC (Clustering Coefficient), EC (Eigenvector Centrality), BC (Betweenness Centrality), and CL (Closeness Centrality).
%=========================================================
\subsection{Statistical validation}

When comparing the properties of two networks, such as retracted and non-retracted collaboration networks, statistical validation is crucial to ensure that observed differences are not due to random variation but represent meaningful structural or behavioral distinctions. Statistical methods, like the $t$-test, help quantify these differences and determine their significance. The $t$-test is a statistical method used to compare the means of two groups (e.g., properties from two networks) and determine whether their differences are significant. It is particularly useful when analyzing network properties because these metrics often follow normal or near-normal distributions. A significant $p$-value (e.g., $p<0.05$) indicates that the difference in the property is unlikely due to chance. We define the hypothesis as
\begin{itemize}
    \item \textbf{Null Hypothesis ($H_0$):} There is no significant difference in the metric between the two networks.
    \item \textbf{Alternative Hypothesis ($H_1$):} There is a significant difference in the metric between the two networks.
\end{itemize}

Table~\ref{tab: T-test} summarizes the results of t-tests conducted on various network metrics to determine significant differences between two groups. Degree Centrality, Average Weighted Degree, Average Path Length, Assortativity, Eigenvector Centrality, and Closeness Centrality show statistically significant differences between the two groups with $p<0.05$. However, metrics like transitivity, clustering coefficient, and betweenness centrality do not show statistically significant differences. These results suggest that certain local and global clustering properties, as well as node bridging roles, are relatively similar between the two groups. Overall, the analysis highlights significant differences in several key metrics, particularly those related to centrality and connectivity, while other metrics remain consistent across the groups. 

%~~~~~~~~~~~~~~~~~~~~  T-test~~~~~~~~~~~~~~~~~~~~~~~~~~~~~~~~~~~~~~~~~~~~~~~~~~~
\begin{table}[!ht]
\centering
\caption{Statistical significance of all network metrics.}
\begin{tabular}{|l|c|c|}
\hline
\textbf{Network Metric}        & \textbf{T-Statistic} & \textbf{p-value} \\ \hline
Degree Centrality      & -3.2052              & \textbf{0.0024}           \\ \hline
Weighted Degree        & 2.2108               & \textbf{0.0322}           \\ \hline
Average Path Length    & 3.2052               & \textbf{0.0024 }          \\ \hline
Assortativity          & 3.3185               & \textbf{0.0016}           \\ \hline
Transitivity           & -0.3035              & 0.7627           \\ \hline
Clustering Coefficient & 1.502                & 0.142            \\ \hline
Eigenvector Centrality & -3.2965              &\textbf{ 0.0018}           \\ \hline
Betweenness Centrality  & -1.9524              & 0.0596           \\ \hline
Closeness Centrality   & -3.0387              &\textbf{ 0.004}            \\ \hline
\end{tabular}
\label{tab: T-test}
\end{table}
%~~~~~~~~~~~~~~~~~~~~~~~~~~~~~~~~~~~~~~~~~~~~~~~~~~~~~~~~~~~~~~~~~~~~~~~
%~~~~~~~~~~ ccdf plot of 3 authors~~~~~~~~~~~~~~~~~~~~~~~~~~~~~~~~~~~~~~~~~~~~~~~~~~~~~~~~~~~~~
\begin{figure}[!ht]
    \centering
\includegraphics[width=0.75\linewidth]{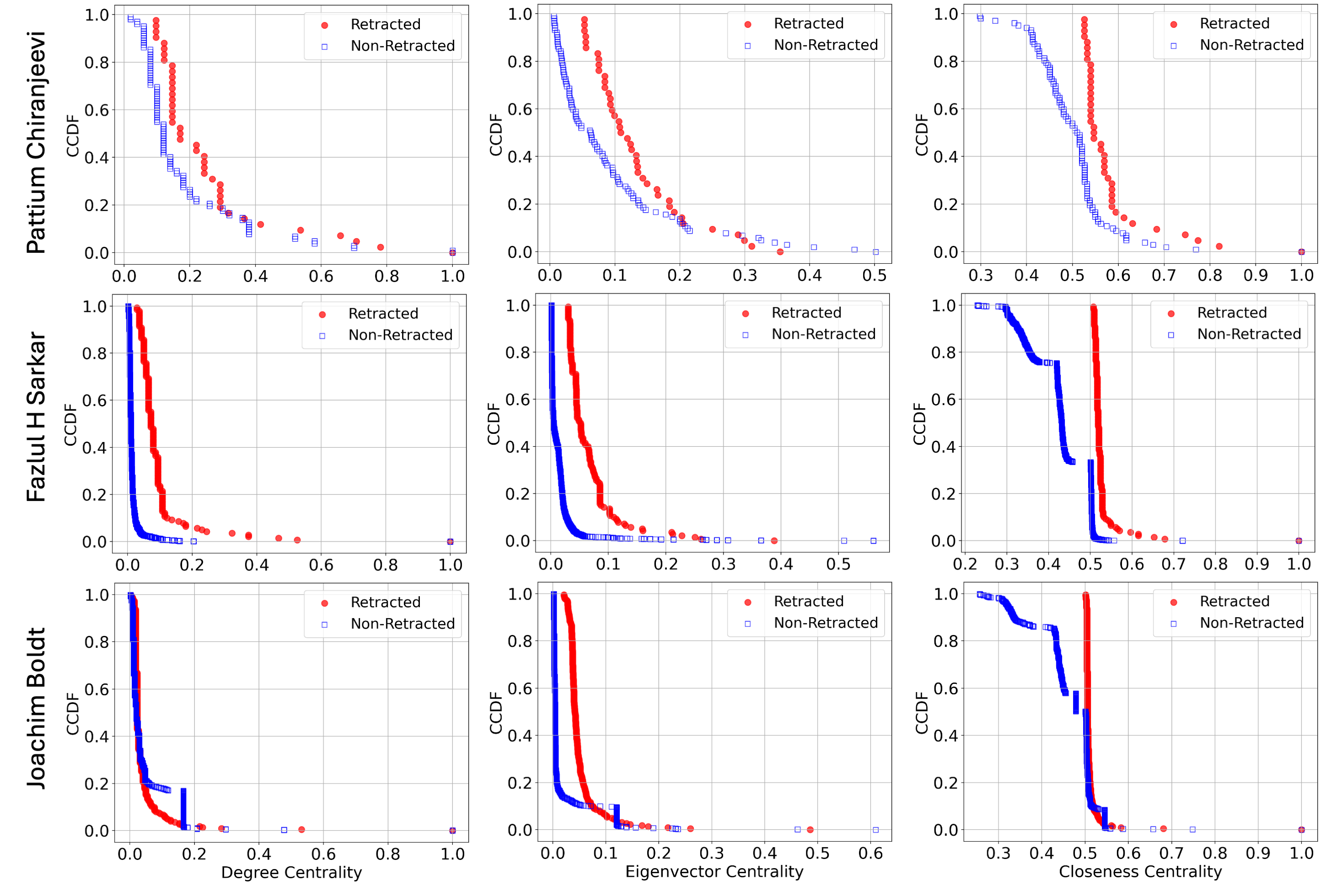} 
\includegraphics[width=0.75\linewidth]{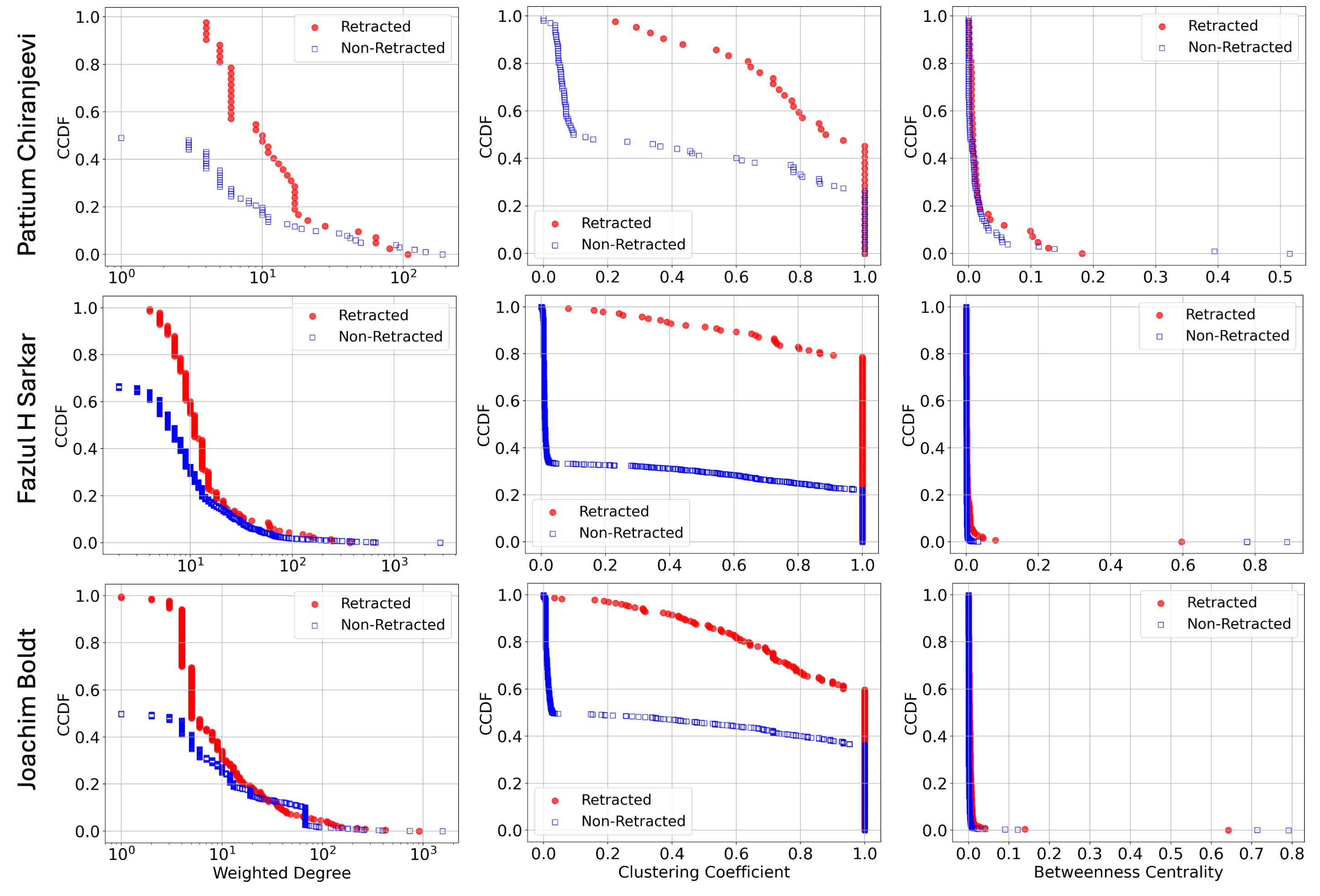} 
\caption{CCDF of author's collaboration network of retracted and non-retracted publications. Upper: (\textit{left}) Degree centrality, (\textit{middle}) Eigenvector centrality, and (\textit{right}) Closeness centrality. Lower: (\textit{left}) Weighted degree, (\textit{middle}) Clustering coefficient, and (\textit{right}) Betweenness centrality.}
\label{fig:CCDF}
\end{figure}
%~~~~~~~~~~~~~~~~~~~~~~~~~~~~~~~~~~~~~~~~~~~~~~~~~~~~~~~~~~~~~~~~~~~~~~~

\subsection{Cohen's $d$ test}
Cohen's $d$ is essential for understanding the practical or real-world significance of differences between groups.While a $t$-test assesses the statistical significance of differences, Cohen's $d$ offers a standardized measure of the effect size, making it a crucial tool for research requiring quantification of the magnitude of these differences. Cohen's $d$ is a standardized measure of the effect size that quantifies the difference between two groups in terms of standard deviation units. Cohen’s $d$ is often paired with a $t$-test to describe the magnitude of the observed effect, adding valuable context to the statistical significance.

A $t$-test determines whether the difference between two groups is statistically significant. However, it does not offer insight into the size or practical importance of the difference. Cohen's $d$ fills this gap by quantifying the effect size. Cohen's $d$ complements the $t$-test by quantifying how large the difference is, allowing for practical interpretation~\citep{cohen1988statistical}.

A small $p$-value in a $t$-test can result from a very large sample size, even if the actual effect is negligible. Cohen's $d$ helps identify whether the observed difference is meaningful or practically significant, irrespective of sample size. Since Cohen's $d$ is standardized, it allows researchers to compare effect sizes across different experiments, populations, or contexts. Cohen's \( d \) is calculated as:
\begin{equation}
  d = \frac{\bar{X}_1 - \bar{X}_2}{s}  
\end{equation}

Where: \[s = \sqrt{\frac{(n_1 - 1)s_1^2 + (n_2 - 1)s_2^2}{n_1 + n_2 - 2}}\]
\begin{itemize}
     \item \( \bar{X}_1, \bar{X}_2 \): Mean values of the two groups  
 \item \( s \): Pooled standard deviation  
 \item \( n_1, n_2 \): Sample sizes of the two groups  
 \item \( s_1, s_2 \): Standard deviations of the two groups  
\end{itemize}

The table~\ref{tab:Cohen} summarizes the effect sizes of various network metrics using Cohen's $d$, indicating the standardized differences between two groups. Metrics such as Assortativity ($d=0.857$), Average Path Length ($d=0.828$), and Eigenvector Centrality ($d=-0.851$) show large effect sizes, highlighting significant differences between the groups. For example, assortativity is higher in one group, suggesting a stronger tendency for nodes to connect with others of similar degree, while the average path length is longer in the same group, indicating less efficient connectivity. On the other hand, eigenvector centrality is significantly lower in one group, reflecting reduced influence of high-degree nodes.

Medium effect sizes are observed for metrics like Average Weighted Degree ($d=0.571$) and Betweenness Centrality ($d=-0.504$), showing moderate differences in connection weights and the role of nodes in bridging different parts of the network. Small to negligible effects are noted for metrics like Clustering Coefficient ($d = 0.388$) and Transitivity ($d=-0.078$), suggesting minimal differences in local clustering or triangle formation between the groups. Negative effect sizes for metrics such as Degree Centrality ($d=-0.828$) and Closeness Centrality ($d=-0.785$) indicate that nodes in one group are generally less connected and less central than in the other. Overall, the analysis reveals significant structural differences in Assortativity, Average Path Length, and Eigenvector Centrality , while others remain relatively similar between the two groups.

%~~~~~~~~~~~~~~~~~~Cohen's d Test~~~~~~~~~~~~~~~~~~~~~~~~~~~~~~~~~~~~~~~~~~~~~~~~~~~~~
\begin{table}[!ht]
\centering
\caption{Cohen's $d$ Test.}
\begin{tabular}{|l|c|l|c|}
\hline
\textbf{Network Metric}         & \textbf{Cohen's $d$} & \textbf{Network Metric}        & \textbf{Cohen's $d$} \\ \hline
Degree Centrality       & \textbf{-0.8275}   & Clustering Coefficient & 0.3878             \\ \hline
Average Weighted Degree & 0.5708             & Eigenvector Centrality & -0.8511            \\ \hline
Average Path Length     & \textbf{0.8275}    & Betweenness Centrality  & -0.5041            \\ \hline
Assortativity           & 0.8568             & Closeness Centrality   & -0.7845            \\ \hline
Transitivity            & -0.0783            & -                      & -                  \\ \hline
\end{tabular}
\label{tab:Cohen}
\end{table}
%=========================================================

%=========================================================

%=========================================================
\section{Conclusion}

This study provides an in-depth analysis of the structural and dynamic distinctions between retracted and non-retracted collaboration networks, offering valuable insights into the influence of network properties on scientific outcomes. By evaluating metrics such as degree centrality, weighted degree, clustering coefficients, and assortativity, the research highlights key differences. For instance, retracted networks exhibit more centralized and hierarchical structures, with high correlations between degree and eigenvector centrality, indicating a reliance on a few influential nodes. Compared to retracted networks, non-retracted ones show greater clustering and more widespread collaborations, reflecting a balanced and robust network structure.

The findings highlight key differences between retracted and non-retracted networks. In retracted networks, high-centrality nodes often play pivotal roles in spreading flawed research. In contrast, non-retracted networks often feature cohesive sub-networks that foster collaboration and encourage innovation. Statistical validation using $t$-tests and Cohen's $d$ confirms these findings, revealing significant differences in metrics such as average path length and eigenvector centrality.

These results highlight the need to promote ethical research practices and balanced collaboration networks to minimize the risk of retractions. Implementing robust institutional policies and monitoring key network metrics can act as early warning systems to detect problematic patterns, offering a practical approach to improving the reliability and integrity of scientific collaborations. Future research could build on this framework by integrating interdisciplinary analyzes and exploring external factors such as funding, policy interventions, and cultural influences on academic collaborations. Gaining a deeper understanding of network dynamics not only helps tackle the issues associated with retractions, but also guides efforts to build a robust and ethical global research ecosystem.

%=========================================================
\subsection{Limitations}
First, the present study focuses on 30 authors with heavy retractions and cannot be representative of broader academic trends. Second, this study focuses on structural metrics without looking at external factors such as funding or institutional policies. Third, we are not able to include some of the authors for incomplete data or insufficient data in the databases.

%%~~~~~~~~~~~~~~~~~~~~~~~~~Retracted metrics~~~~~~~~~~~~~~~~~~~~~~~~~~~~~~~~
\begin{table}[!ht]
\centering
\caption{Network properties of retracted papers author's collaboration network: Network Diameter (ND), Average Degree Centrality (DC), Average Weighted Degree (WD), Average Path Length (APL), Average Assortativity, Average Transitivity, Average Clustering Coefficient (CC), Average Eigenvector Centrality (EC), Average Betweenness Centrality (BC), Average Closeness Centrality (CL).}
 \resizebox{\textwidth}{!}{ 
\begin{tabular}{|c|l|c|ccccccccc|}
\hline
\multicolumn{1}{|l|}{\multirow{2}{*}{\textbf{S.No}}} & \multirow{2}{*}{\textbf{Author Name}} & \multicolumn{1}{l|}{\multirow{2}{*}{\textbf{ND}}} & \multicolumn{9}{c|}{\textbf{Average}}             \\ \cline{4-12} 
\multicolumn{1}{|l|}{}                               &                                       & \multicolumn{1}{l|}{}                             & \multicolumn{1}{c|}{\textbf{DC}} & \multicolumn{1}{c|}{\textbf{WD}} & \multicolumn{1}{c|}{\textbf{PL}} & \multicolumn{1}{c|}{\textbf{Assortativity}} & \multicolumn{1}{c|}{\textbf{Transitivity}} & \multicolumn{1}{c|}{\textbf{CC}} & \multicolumn{1}{c|}{\textbf{EC}} & \multicolumn{1}{c|}{\textbf{BC}} & \textbf{CoC} \\ \hline
1                                                    & Ali Nazari                            & 2                                                 & \multicolumn{1}{c|}{0.1496}      & \multicolumn{1}{c|}{5.0857}      & \multicolumn{1}{c|}{1.8504}      & \multicolumn{1}{c|}{-0.3339}                & \multicolumn{1}{c|}{0.3326}                & \multicolumn{1}{c|}{0.8035}      & \multicolumn{1}{c|}{0.146}       & \multicolumn{1}{c|}{0.0258}      & 0.5473       \\ \hline
2                                                    & Annarosa Leri                         & 2                                                 & \multicolumn{1}{c|}{0.1852}      & \multicolumn{1}{c|}{27.4094}     & \multicolumn{1}{c|}{1.8148}      & \multicolumn{1}{c|}{-0.2171}                & \multicolumn{1}{c|}{0.4927}                & \multicolumn{1}{c|}{0.8848}      & \multicolumn{1}{c|}{0.0686}      & \multicolumn{1}{c|}{0.0055}      & 0.5563       \\ \hline
3                                                    & Ashok Pandey                          & 2                                                 & \multicolumn{1}{c|}{0.0756}      & \multicolumn{1}{c|}{8.1651}      & \multicolumn{1}{c|}{1.9244}      & \multicolumn{1}{c|}{-0.2201}                & \multicolumn{1}{c|}{0.2806}                & \multicolumn{1}{c|}{0.9144}      & \multicolumn{1}{c|}{0.0759}      & \multicolumn{1}{c|}{0.0086}      & 0.5221       \\ \hline
4                                                    & Diederik A. Stapel                    & 2                                                 & \multicolumn{1}{c|}{0.081}       & \multicolumn{1}{c|}{3.0769}      & \multicolumn{1}{c|}{1.919}       & \multicolumn{1}{c|}{-0.5108}                & \multicolumn{1}{c|}{0.1123}                & \multicolumn{1}{c|}{0.5606}      & \multicolumn{1}{c|}{0.1333}      & \multicolumn{1}{c|}{0.0248}      & 0.5272       \\ \hline
5                                                    & Dong Mei Wu                           & 2                                                 & \multicolumn{1}{c|}{0.2424}      & \multicolumn{1}{c|}{18.6667}     & \multicolumn{1}{c|}{1.7576}      & \multicolumn{1}{c|}{-0.4684}                & \multicolumn{1}{c|}{0.4482}                & \multicolumn{1}{c|}{0.8721}      & \multicolumn{1}{c|}{0.0971}      & \multicolumn{1}{c|}{0.01}        & 0.5816       \\ \hline
6                                                    & Ekta Roy                              & 2                                                 & \multicolumn{1}{c|}{0.6282}      & \multicolumn{1}{c|}{7.5385}      & \multicolumn{1}{c|}{1.3718}      & \multicolumn{1}{c|}{-0.5503}                & \multicolumn{1}{c|}{0.6857}                & \multicolumn{1}{c|}{0.8501}      & \multicolumn{1}{c|}{0.2671}      & \multicolumn{1}{c|}{0.0338}      & 0.7603       \\ \hline
7                                                    & Fazlul H Sarkar                       & 2                                                 & \multicolumn{1}{c|}{0.0976}      & \multicolumn{1}{c|}{13.5714}     & \multicolumn{1}{c|}{1.9024}      & \multicolumn{1}{c|}{-0.2091}                & \multicolumn{1}{c|}{0.3635}                & \multicolumn{1}{c|}{0.9083}      & \multicolumn{1}{c|}{0.068}       & \multicolumn{1}{c|}{0.0065}      & 0.5282       \\ \hline
8                                                    & Gunter G Hempelmann                   & 2                                                 & \multicolumn{1}{c|}{0.0718}      & \multicolumn{1}{c|}{8.4706}      & \multicolumn{1}{c|}{1.9282}      & \multicolumn{1}{c|}{-0.3629}                & \multicolumn{1}{c|}{0.1507}                & \multicolumn{1}{c|}{0.8619}      & \multicolumn{1}{c|}{0.075}       & \multicolumn{1}{c|}{0.0079}      & 0.5227       \\ \hline
9                                                    & Hans J Eysenck                        & 2                                                 & \multicolumn{1}{c|}{0.113}       & \multicolumn{1}{c|}{4.7442}      & \multicolumn{1}{c|}{1.887}       & \multicolumn{1}{c|}{-0.4433}                & \multicolumn{1}{c|}{0.3375}                & \multicolumn{1}{c|}{0.6636}      & \multicolumn{1}{c|}{0.1126}      & \multicolumn{1}{c|}{0.0216}      & 0.5362       \\ \hline
10                                                   & Hironobu Ueshima                      & 2                                                 & \multicolumn{1}{c|}{0.1402}      & \multicolumn{1}{c|}{4.4848}      & \multicolumn{1}{c|}{1.8598}      & \multicolumn{1}{c|}{-0.7098}                & \multicolumn{1}{c|}{0.1636}                & \multicolumn{1}{c|}{0.8923}      & \multicolumn{1}{c|}{0.148}       & \multicolumn{1}{c|}{0.0277}      & 0.55         \\ \hline
11                                                   & Hiroyoshi Tanaka                      & 2                                                 & \multicolumn{1}{c|}{0.3676}      & \multicolumn{1}{c|}{5.8824}      & \multicolumn{1}{c|}{1.6324}      & \multicolumn{1}{c|}{-0.5933}                & \multicolumn{1}{c|}{0.5103}                & \multicolumn{1}{c|}{0.8446}      & \multicolumn{1}{c|}{0.2195}      & \multicolumn{1}{c|}{0.0422}      & 0.6324       \\ \hline
12                                                   & Hyung-In Moon                         & 2                                                 & \multicolumn{1}{c|}{0.1144}      & \multicolumn{1}{c|}{5.375}       & \multicolumn{1}{c|}{1.8856}      & \multicolumn{1}{c|}{-0.2977}                & \multicolumn{1}{c|}{0.2932}                & \multicolumn{1}{c|}{0.8349}      & \multicolumn{1}{c|}{0.1212}      & \multicolumn{1}{c|}{0.0193}      & 0.5354       \\ \hline
13                                                   & Jan Hendrik Schon                     & 2                                                 & \multicolumn{1}{c|}{0.2154}      & \multicolumn{1}{c|}{5.3846}      & \multicolumn{1}{c|}{1.7846}      & \multicolumn{1}{c|}{-0.3445}                & \multicolumn{1}{c|}{0.4415}                & \multicolumn{1}{c|}{0.9062}      & \multicolumn{1}{c|}{0.1761}      & \multicolumn{1}{c|}{0.0327}      & 0.5692       \\ \hline
14                                                   & Joachim Boldt                         & 2                                                 & \multicolumn{1}{c|}{0.0427}      & \multicolumn{1}{c|}{9.4798}      & \multicolumn{1}{c|}{1.9573}      & \multicolumn{1}{c|}{-0.2664}                & \multicolumn{1}{c|}{0.1454}                & \multicolumn{1}{c|}{0.8358}      & \multicolumn{1}{c|}{0.0526}      & \multicolumn{1}{c|}{0.0043}      & 0.5124       \\ \hline
15                                                   & Jose Luis Calvo-Guirado               & 2                                                 & \multicolumn{1}{c|}{0.1016}      & \multicolumn{1}{c|}{9.9596}      & \multicolumn{1}{c|}{1.8984}      & \multicolumn{1}{c|}{-0.2051}                & \multicolumn{1}{c|}{0.3252}                & \multicolumn{1}{c|}{0.8645}      & \multicolumn{1}{c|}{0.0841}      & \multicolumn{1}{c|}{0.0093}      & 0.5296       \\ \hline
16                                                   & Jun Iwamoto                           & 2                                                 & \multicolumn{1}{c|}{0.2685}      & \multicolumn{1}{c|}{7.5172}      & \multicolumn{1}{c|}{1.7315}      & \multicolumn{1}{c|}{-0.4754}                & \multicolumn{1}{c|}{0.4218}                & \multicolumn{1}{c|}{0.8393}      & \multicolumn{1}{c|}{0.1656}      & \multicolumn{1}{c|}{0.0271}      & 0.5914       \\ \hline
17                                                   & Li Zhang                              & 2                                                 & \multicolumn{1}{c|}{0.027}       & \multicolumn{1}{c|}{9.2986}      & \multicolumn{1}{c|}{1.973}       & \multicolumn{1}{c|}{-0.1392}                & \multicolumn{1}{c|}{0.2186}                & \multicolumn{1}{c|}{0.9644}      & \multicolumn{1}{c|}{0.0396}      & \multicolumn{1}{c|}{0.0028}      & 0.5076       \\ \hline
18                                                   & Naoki Mori                            & 2                                                 & \multicolumn{1}{c|}{0.2115}      & \multicolumn{1}{c|}{18.8222}     & \multicolumn{1}{c|}{1.7885}      & \multicolumn{1}{c|}{-0.201}                 & \multicolumn{1}{c|}{0.5145}                & \multicolumn{1}{c|}{0.8045}      & \multicolumn{1}{c|}{0.0882}      & \multicolumn{1}{c|}{0.009}       & 0.5646       \\ \hline
19                                                   & Pattium Chiranjeevi                   & 2                                                 & \multicolumn{1}{c|}{0.2602}      & \multicolumn{1}{c|}{10.6667}     & \multicolumn{1}{c|}{1.7398}      & \multicolumn{1}{c|}{-0.3463}                & \multicolumn{1}{c|}{0.4688}                & \multicolumn{1}{c|}{0.817}       & \multicolumn{1}{c|}{0.135}       & \multicolumn{1}{c|}{0.0185}      & 0.5851       \\ \hline
20                                                   & Prashant K Sharma                     & 2                                                 & \multicolumn{1}{c|}{0.3684}      & \multicolumn{1}{c|}{7}           & \multicolumn{1}{c|}{1.6316}      & \multicolumn{1}{c|}{-0.3139}                & \multicolumn{1}{c|}{0.6105}                & \multicolumn{1}{c|}{0.8654}      & \multicolumn{1}{c|}{0.1978}      & \multicolumn{1}{c|}{0.0351}      & 0.6256       \\ \hline
21                                                   & Saber Khelaifia                       & 2                                                 & \multicolumn{1}{c|}{0.1925}      & \multicolumn{1}{c|}{17.1333}     & \multicolumn{1}{c|}{1.8075}      & \multicolumn{1}{c|}{-0.2813}                & \multicolumn{1}{c|}{0.4666}                & \multicolumn{1}{c|}{0.8294}      & \multicolumn{1}{c|}{0.0904}      & \multicolumn{1}{c|}{0.0092}      & 0.5592       \\ \hline
22                                                   & Shahaboddin Shamshirband              & 2                                                 & \multicolumn{1}{c|}{0.0778}      & \multicolumn{1}{c|}{8.243}       & \multicolumn{1}{c|}{1.9222}      & \multicolumn{1}{c|}{-0.2395}                & \multicolumn{1}{c|}{0.2468}                & \multicolumn{1}{c|}{0.8799}      & \multicolumn{1}{c|}{0.0781}      & \multicolumn{1}{c|}{0.0088}      & 0.5229       \\ \hline
23                                                   & Shigeaki Kato                         & 2                                                 & \multicolumn{1}{c|}{0.1227}      & \multicolumn{1}{c|}{19.8773}     & \multicolumn{1}{c|}{1.8773}      & \multicolumn{1}{c|}{-0.212}                 & \multicolumn{1}{c|}{0.4005}                & \multicolumn{1}{c|}{0.8565}      & \multicolumn{1}{c|}{0.064}       & \multicolumn{1}{c|}{0.0054}      & 0.5358       \\ \hline
24                                                   & Soon-Gi Shin                          & 2                                                 & \multicolumn{1}{c|}{0.5}         & \multicolumn{1}{c|}{1.5}         & \multicolumn{1}{c|}{1.5}         & \multicolumn{1}{c|}{-1}                     & \multicolumn{1}{c|}{0}                     & \multicolumn{1}{c|}{0}           & \multicolumn{1}{c|}{0.483}       & \multicolumn{1}{c|}{0.25}        & 0.7          \\ \hline
25                                                   & Stefan W. Suttner                     & 2                                                 & \multicolumn{1}{c|}{0.2041}      & \multicolumn{1}{c|}{9.7959}      & \multicolumn{1}{c|}{1.7959}      & \multicolumn{1}{c|}{-0.4964}                & \multicolumn{1}{c|}{0.3235}                & \multicolumn{1}{c|}{0.8282}      & \multicolumn{1}{c|}{0.1244}      & \multicolumn{1}{c|}{0.0169}      & 0.5687       \\ \hline
26                                                   & Victor Grech                          & 2                                                 & \multicolumn{1}{c|}{0.1378}      & \multicolumn{1}{c|}{4.6857}      & \multicolumn{1}{c|}{1.8622}      & \multicolumn{1}{c|}{-0.384}                 & \multicolumn{1}{c|}{0.3095}                & \multicolumn{1}{c|}{0.8062}      & \multicolumn{1}{c|}{0.1397}      & \multicolumn{1}{c|}{0.0261}      & 0.5441       \\ \hline
27                                                   & Yogeshwer Shukla                      & 2                                                 & \multicolumn{1}{c|}{0.3251}      & \multicolumn{1}{c|}{9.1034}      & \multicolumn{1}{c|}{1.6749}      & \multicolumn{1}{c|}{-0.207}                 & \multicolumn{1}{c|}{0.5932}                & \multicolumn{1}{c|}{0.8384}      & \multicolumn{1}{c|}{0.1686}      & \multicolumn{1}{c|}{0.025}       & 0.6057       \\ \hline
28                                                   & Yoshihiro Sato                        & 2                                                 & \multicolumn{1}{c|}{0.1419}      & \multicolumn{1}{c|}{8.9375}      & \multicolumn{1}{c|}{1.8581}      & \multicolumn{1}{c|}{-0.3174}                & \multicolumn{1}{c|}{0.3333}                & \multicolumn{1}{c|}{0.8381}      & \multicolumn{1}{c|}{0.1066}      & \multicolumn{1}{c|}{0.0138}      & 0.5431       \\ \hline
29                                                   & Yoshitaka Fujii                       & 2                                                 & \multicolumn{1}{c|}{0.154}       & \multicolumn{1}{c|}{5.3889}      & \multicolumn{1}{c|}{1.846}       & \multicolumn{1}{c|}{-0.4682}                & \multicolumn{1}{c|}{0.3021}                & \multicolumn{1}{c|}{0.7392}      & \multicolumn{1}{c|}{0.1395}      & \multicolumn{1}{c|}{0.0249}      & 0.5496       \\ \hline
30                                                   & Yuhji Saitoh                          & 2                                                 & \multicolumn{1}{c|}{0.2197}      & \multicolumn{1}{c|}{7.0303}      & \multicolumn{1}{c|}{1.7803}      & \multicolumn{1}{c|}{-0.2792}                & \multicolumn{1}{c|}{0.443}                 & \multicolumn{1}{c|}{0.8143}      & \multicolumn{1}{c|}{0.1564}      & \multicolumn{1}{c|}{0.0252}      & 0.5691       \\ \hline
\end{tabular}
}
\label{table:RetProperty}
\end{table}
%%~~~~~~~~~~~~~~~~~~~~~~~~~~~~~~~~~~~~~~~~~~~~~~

%%~~~~~~~~~~~~~~~~~~Non_retracted metrics~~~~~~~~~~~~~~~~~~~~~~~~~~~~~~~~~~~~~~~~~~~~
\begin{table}[!ht]
\centering
\caption{Network properties of non-retracted author's collaboration network: Network Diameter (ND), Average Degree Centrality (DC), Average Weighted Degree (WD), Average Path Length (APL), Average Assortativity, AverageTransitivity, Average Clustering Coefficient (CC), Average Eigenvector Centrality (EC), Average Betweenness Centrality (BC), Average Closeness Centrality (CL).}
 \resizebox{\textwidth}{!}{ 
\begin{tabular}{|c|l|c|ccccccccc|}
\hline
\multicolumn{1}{|l|}{\multirow{2}{*}{\textbf{S.No}}} & \multirow{2}{*}{\textbf{Author Name}} & \multicolumn{1}{l|}{\multirow{2}{*}{\textbf{ND}}} & \multicolumn{9}{c|}{\textbf{Average}}  \\ \cline{4-12} 
\multicolumn{1}{|l|}{}                               &                                       & \multicolumn{1}{l|}{}                             & \multicolumn{1}{c|}{\textbf{DC}} & \multicolumn{1}{c|}{\textbf{WD}} & \multicolumn{1}{c|}{\textbf{PL}} & \multicolumn{1}{c|}{\textbf{Assortativity}} & \multicolumn{1}{c|}{\textbf{Transitivity}} & \multicolumn{1}{c|}{\textbf{CC}} & \multicolumn{1}{c|}{\textbf{EC}} & \multicolumn{1}{c|}{\textbf{BC}} & \textbf{CL} \\ \hline
1                                                    & Ali Nazari                            & 2                                                 & \multicolumn{1}{c|}{0.0644}      & \multicolumn{1}{c|}{6.1856}      & \multicolumn{1}{c|}{1.9356}      & \multicolumn{1}{c|}{-0.3168}                & \multicolumn{1}{c|}{0.1756}                & \multicolumn{1}{c|}{0.8784}      & \multicolumn{1}{c|}{0.0817}      & \multicolumn{1}{c|}{0.0098}      & 0.5195      \\ \hline
2                                                    & Annarosa Leri                         & 2                                                 & \multicolumn{1}{c|}{0.0637}      & \multicolumn{1}{c|}{27.711}      & \multicolumn{1}{c|}{1.9363}      & \multicolumn{1}{c|}{-0.2115}                & \multicolumn{1}{c|}{0.2631}                & \multicolumn{1}{c|}{0.8486}      & \multicolumn{1}{c|}{0.0371}      & \multicolumn{1}{c|}{0.0022}      & 0.5182      \\ \hline
3                                                    & Ashok Pandey                          & 2                                                 & \multicolumn{1}{c|}{0.0114}      & \multicolumn{1}{c|}{12.3546}     & \multicolumn{1}{c|}{1.9886}      & \multicolumn{1}{c|}{-0.1362}                & \multicolumn{1}{c|}{0.0802}                & \multicolumn{1}{c|}{0.8753}      & \multicolumn{1}{c|}{0.0216}      & \multicolumn{1}{c|}{0.0009}      & 0.5031      \\ \hline
4                                                    & Diederik A. Stapel                    & 2                                                 & \multicolumn{1}{c|}{0.063}       & \multicolumn{1}{c|}{3.9048}      & \multicolumn{1}{c|}{1.937}       & \multicolumn{1}{c|}{-0.3821}                & \multicolumn{1}{c|}{0.1329}                & \multicolumn{1}{c|}{0.7541}      & \multicolumn{1}{c|}{0.1031}      & \multicolumn{1}{c|}{0.0154}      & 0.5201      \\ \hline
5                                                    & Dong Mei Wu                           & 2                                                 & \multicolumn{1}{c|}{0.1307}      & \multicolumn{1}{c|}{17.5111}     & \multicolumn{1}{c|}{1.8693}      & \multicolumn{1}{c|}{-0.307}                 & \multicolumn{1}{c|}{0.3791}                & \multicolumn{1}{c|}{0.8863}      & \multicolumn{1}{c|}{0.067}       & \multicolumn{1}{c|}{0.0065}      & 0.5393      \\ \hline
6                                                    & Ekta Roy                              & 2                                                 & \multicolumn{1}{c|}{0.2826}      & \multicolumn{1}{c|}{6.5}         & \multicolumn{1}{c|}{1.7174}      & \multicolumn{1}{c|}{-0.2436}                & \multicolumn{1}{c|}{0.5691}                & \multicolumn{1}{c|}{0.8773}      & \multicolumn{1}{c|}{0.1896}      & \multicolumn{1}{c|}{0.0326}      & 0.5906      \\ \hline
7                                                    & Fazul H Sarkar                        & 2                                                 & \multicolumn{1}{c|}{0.0154}      & \multicolumn{1}{c|}{13.2987}     & \multicolumn{1}{c|}{1.9846}      & \multicolumn{1}{c|}{-0.1263}                & \multicolumn{1}{c|}{0.1053}                & \multicolumn{1}{c|}{0.8701}      & \multicolumn{1}{c|}{0.025}       & \multicolumn{1}{c|}{0.0011}      & 0.5042      \\ \hline
8                                                    & Gunter G Hempelmann                   & 2                                                 & \multicolumn{1}{c|}{0.0253}      & \multicolumn{1}{c|}{17.8872}     & \multicolumn{1}{c|}{1.9747}      & \multicolumn{1}{c|}{-0.1295}                & \multicolumn{1}{c|}{0.3885}                & \multicolumn{1}{c|}{0.8629}      & \multicolumn{1}{c|}{0.0148}      & \multicolumn{1}{c|}{0.0014}      & 0.5069      \\ \hline
9                                                    & Hans J Eysenck                        & 2                                                 & \multicolumn{1}{c|}{0.0368}      & \multicolumn{1}{c|}{4.4878}      & \multicolumn{1}{c|}{1.9632}      & \multicolumn{1}{c|}{-0.3435}                & \multicolumn{1}{c|}{0.1062}                & \multicolumn{1}{c|}{0.7235}      & \multicolumn{1}{c|}{0.0706}      & \multicolumn{1}{c|}{0.008}       & 0.5114      \\ \hline
10                                                   & Hironobu Ueshima                      & 2                                                 & \multicolumn{1}{c|}{0.1554}      & \multicolumn{1}{c|}{6.6818}      & \multicolumn{1}{c|}{1.8446}      & \multicolumn{1}{c|}{-0.2519}                & \multicolumn{1}{c|}{0.4327}                & \multicolumn{1}{c|}{0.8863}      & \multicolumn{1}{c|}{0.1306}      & \multicolumn{1}{c|}{0.0201}      & 0.5476      \\ \hline
11                                                   & Hiroyoshi Tanaka                      & 2                                                 & \multicolumn{1}{c|}{0.2737}      & \multicolumn{1}{c|}{5.2}         & \multicolumn{1}{c|}{1.7263}      & \multicolumn{1}{c|}{-0.6472}                & \multicolumn{1}{c|}{0.3843}                & \multicolumn{1}{c|}{0.8793}      & \multicolumn{1}{c|}{0.198}       & \multicolumn{1}{c|}{0.0404}      & 0.5972      \\ \hline
12                                                   & Hyung-In Moon                         & 2                                                 & \multicolumn{1}{c|}{0.0705}      & \multicolumn{1}{c|}{10.3649}     & \multicolumn{1}{c|}{1.9295}      & \multicolumn{1}{c|}{-0.1692}                & \multicolumn{1}{c|}{0.3373}                & \multicolumn{1}{c|}{0.8756}      & \multicolumn{1}{c|}{0.0672}      & \multicolumn{1}{c|}{0.0064}      & 0.5201      \\ \hline
13                                                   & Jan Hendrik Schon                     & 2                                                 & \multicolumn{1}{c|}{0.1092}      & \multicolumn{1}{c|}{8.08}        & \multicolumn{1}{c|}{1.8908}      & \multicolumn{1}{c|}{-0.2959}                & \multicolumn{1}{c|}{0.2889}                & \multicolumn{1}{c|}{0.8714}      & \multicolumn{1}{c|}{0.0965}      & \multicolumn{1}{c|}{0.0122}      & 0.5329      \\ \hline
14                                                   & Joachim Boldt                         & 2                                                 & \multicolumn{1}{c|}{0.0504}      & \multicolumn{1}{c|}{20.3506}     & \multicolumn{1}{c|}{1.9496}      & \multicolumn{1}{c|}{-0.1511}                & \multicolumn{1}{c|}{0.5996}                & \multicolumn{1}{c|}{0.8952}      & \multicolumn{1}{c|}{0.0235}      & \multicolumn{1}{c|}{0.0024}      & 0.514       \\ \hline
15                                                   & Jose L Calvo-Guirado                  & 2                                                 & \multicolumn{1}{c|}{0.0299}      & \multicolumn{1}{c|}{10.5915}     & \multicolumn{1}{c|}{1.9701}      & \multicolumn{1}{c|}{-0.1753}                & \multicolumn{1}{c|}{0.1508}                & \multicolumn{1}{c|}{0.879}       & \multicolumn{1}{c|}{0.0388}      & \multicolumn{1}{c|}{0.0027}      & 0.5084      \\ \hline
16                                                   & Jun Iwamoto                           & 2                                                 & \multicolumn{1}{c|}{0.0929}      & \multicolumn{1}{c|}{8.6383}      & \multicolumn{1}{c|}{1.9071}      & \multicolumn{1}{c|}{-0.2687}                & \multicolumn{1}{c|}{0.3082}                & \multicolumn{1}{c|}{0.894}       & \multicolumn{1}{c|}{0.0853}      & \multicolumn{1}{c|}{0.0099}      & 0.5275      \\ \hline
17                                                   & Li Zhang                              & 2                                                 & \multicolumn{1}{c|}{0.0284}      & \multicolumn{1}{c|}{38.4065}     & \multicolumn{1}{c|}{1.9716}      & \multicolumn{1}{c|}{-0.0808}                & \multicolumn{1}{c|}{0.352}                 & \multicolumn{1}{c|}{0.8907}      & \multicolumn{1}{c|}{0.0186}      & \multicolumn{1}{c|}{0.0007}      & 0.5075      \\ \hline
18                                                   & Naoki Mori                            & 2                                                 & \multicolumn{1}{c|}{0.0333}      & \multicolumn{1}{c|}{16.9194}     & \multicolumn{1}{c|}{1.9667}      & \multicolumn{1}{c|}{-0.1118}                & \multicolumn{1}{c|}{0.2429}                & \multicolumn{1}{c|}{0.8644}      & \multicolumn{1}{c|}{0.0338}      & \multicolumn{1}{c|}{0.0019}      & 0.509       \\ \hline
19                                                   & Pattium Chiranjeevi                   & 2                                                 & \multicolumn{1}{c|}{0.1875}      & \multicolumn{1}{c|}{9.3725}      & \multicolumn{1}{c|}{1.8125}      & \multicolumn{1}{c|}{-0.3907}                & \multicolumn{1}{c|}{0.3946}                & \multicolumn{1}{c|}{0.8222}      & \multicolumn{1}{c|}{0.1161}      & \multicolumn{1}{c|}{0.0166}      & 0.5596      \\ \hline
20                                                   & Prashant K Sharma                     & 2                                                 & \multicolumn{1}{c|}{0.0653}      & \multicolumn{1}{c|}{7.5726}      & \multicolumn{1}{c|}{1.9347}      & \multicolumn{1}{c|}{-0.2402}                & \multicolumn{1}{c|}{0.2204}                & \multicolumn{1}{c|}{0.899}       & \multicolumn{1}{c|}{0.0758}      & \multicolumn{1}{c|}{0.0081}      & 0.5192      \\ \hline
21                                                   & Saber Khelaifia                       & 2                                                 & \multicolumn{1}{c|}{0.1234}      & \multicolumn{1}{c|}{29.8601}     & \multicolumn{1}{c|}{1.8766}      & \multicolumn{1}{c|}{-0.1881}                & \multicolumn{1}{c|}{0.6212}                & \multicolumn{1}{c|}{0.887}       & \multicolumn{1}{c|}{0.0468}      & \multicolumn{1}{c|}{0.0036}      & 0.5357      \\ \hline
22                                                   & Shahaboddin Shamshirband              & 2                                                 & \multicolumn{1}{c|}{0.0087}      & \multicolumn{1}{c|}{10.078}      & \multicolumn{1}{c|}{1.9913}      & \multicolumn{1}{c|}{-0.1595}                & \multicolumn{1}{c|}{0.0484}                & \multicolumn{1}{c|}{0.8698}      & \multicolumn{1}{c|}{0.0219}      & \multicolumn{1}{c|}{0.0009}      & 0.5024      \\ \hline
23                                                   & Shigeaki Kato                         & 2                                                 & \multicolumn{1}{c|}{0.01}        & \multicolumn{1}{c|}{17.3931}     & \multicolumn{1}{c|}{1.99}        & \multicolumn{1}{c|}{-0.0819}                & \multicolumn{1}{c|}{0.1167}                & \multicolumn{1}{c|}{0.8963}      & \multicolumn{1}{c|}{0.0171}      & \multicolumn{1}{c|}{0.0006}      & 0.5027      \\ \hline
24                                                   & Soon-Gi Shin                          & 2                                                 & \multicolumn{1}{c|}{0.2492}      & \multicolumn{1}{c|}{6.2308}      & \multicolumn{1}{c|}{1.7508}      & \multicolumn{1}{c|}{-0.4924}                & \multicolumn{1}{c|}{0.4249}                & \multicolumn{1}{c|}{0.7276}      & \multicolumn{1}{c|}{0.1714}      & \multicolumn{1}{c|}{0.0313}      & 0.5847      \\ \hline
25                                                   & Stefan W. Suttner                     & 2                                                 & \multicolumn{1}{c|}{0.2859}      & \multicolumn{1}{c|}{38.6029}     & \multicolumn{1}{c|}{1.7141}      & \multicolumn{1}{c|}{0.1352}                 & \multicolumn{1}{c|}{0.9227}                & \multicolumn{1}{c|}{0.9279}      & \multicolumn{1}{c|}{0.0625}      & \multicolumn{1}{c|}{0.0053}      & 0.5949      \\ \hline
26                                                   & Victor Grech                          & 2                                                 & \multicolumn{1}{c|}{0.1332}      & \multicolumn{1}{c|}{38.774}      & \multicolumn{1}{c|}{1.8668}      & \multicolumn{1}{c|}{0.0396}                 & \multicolumn{1}{c|}{0.9276}                & \multicolumn{1}{c|}{0.8955}      & \multicolumn{1}{c|}{0.035}       & \multicolumn{1}{c|}{0.003}       & 0.5402      \\ \hline
27                                                   & Yogeshwer Shukla                      & 2                                                 & \multicolumn{1}{c|}{0.053}       & \multicolumn{1}{c|}{10.7094}     & \multicolumn{1}{c|}{1.947}       & \multicolumn{1}{c|}{-0.143}                 & \multicolumn{1}{c|}{0.3135}                & \multicolumn{1}{c|}{0.8706}      & \multicolumn{1}{c|}{0.0514}      & \multicolumn{1}{c|}{0.0047}      & 0.5149      \\ \hline
28                                                   & Yoshihiro Sato                        & 2                                                 & \multicolumn{1}{c|}{0.0796}      & \multicolumn{1}{c|}{9.7903}      & \multicolumn{1}{c|}{1.9204}      & \multicolumn{1}{c|}{-0.1954}                & \multicolumn{1}{c|}{0.3091}                & \multicolumn{1}{c|}{0.8745}      & \multicolumn{1}{c|}{0.0741}      & \multicolumn{1}{c|}{0.0075}      & 0.5229      \\ \hline
29                                                   & Yoshitaka Fujii                       & 2                                                 & \multicolumn{1}{c|}{0.1423}      & \multicolumn{1}{c|}{5.55}        & \multicolumn{1}{c|}{1.8577}      & \multicolumn{1}{c|}{-0.4485}                & \multicolumn{1}{c|}{0.2946}                & \multicolumn{1}{c|}{0.7028}      & \multicolumn{1}{c|}{0.1317}      & \multicolumn{1}{c|}{0.0226}      & 0.5455      \\ \hline
30                                                   & Yuhji Saitoh                          & 2                                                 & \multicolumn{1}{c|}{0.1707}      & \multicolumn{1}{c|}{7.5111}      & \multicolumn{1}{c|}{1.8293}      & \multicolumn{1}{c|}{-0.2412}                & \multicolumn{1}{c|}{0.4005}                & \multicolumn{1}{c|}{0.8046}      & \multicolumn{1}{c|}{0.1317}      & \multicolumn{1}{c|}{0.0193}      & 0.5521      \\ \hline
\end{tabular}
\label{table:nonretProperty}
 }
\end{table}
%%~~~~~~~~~~~~~~~~~~~~~~~~~~~~~~~~~~~~~~~~~~~~~~

%=========================================================
%\section*{Acknowledgment}
%Dr. Kiran Sharma and Dr. Ziya Uddin appreciate the invaluable learning support provided by the Center for Teaching, Learning, and Development of BML Munjal University. Special thanks are extended to the Research and Development Cell for their financial support through the seed grant (No: BMU/RDC/SG/2024-06), which made this research possible.

\section*{Conflict of interest}
All authors declare no conflict of interest.

\section*{Data availability}
The data utilized in this study are available for reproducibility upon request of the corresponding author.

%=========================================================
\appendix

\bibliographystyle{cas-model2-names}
\bibliography{cas-refs}

\begin{thebibliography}{23}
\expandafter\ifx\csname natexlab\endcsname\relax\def\natexlab#1{#1}\fi
\providecommand{\url}[1]{\texttt{#1}}
\providecommand{\href}[2]{#2}
\providecommand{\path}[1]{#1}
\providecommand{\DOIprefix}{doi:}
\providecommand{\ArXivprefix}{arXiv:}
\providecommand{\URLprefix}{URL: }
\providecommand{\Pubmedprefix}{pmid:}
\providecommand{\doi}[1]{\href{http://dx.doi.org/#1}{\path{#1}}}
\providecommand{\Pubmed}[1]{\href{pmid:#1}{\path{#1}}}
\providecommand{\bibinfo}[2]{#2}
\ifx\xfnm\relax \def\xfnm[#1]{\unskip,\space#1}\fi
%Type = Article
\bibitem[{Azoulay et~al.(2015)Azoulay, Bonatti and Krieger}]{Azoulay2015The}
\bibinfo{author}{Azoulay, P.}, \bibinfo{author}{Bonatti, A.}, \bibinfo{author}{Krieger, J.L.}, \bibinfo{year}{2015}.
\newblock \bibinfo{title}{The career effects of scandal: Evidence from scientific retractions}.
\newblock \bibinfo{journal}{Sociology of Innovation eJournal} \DOIprefix\doi{10.1016/J.RESPOL.2017.07.003}.
%Type = Book
\bibitem[{Barabasi(2016)}]{barabasi2016network}
\bibinfo{author}{Barabasi, A.L.}, \bibinfo{year}{2016}.
\newblock \bibinfo{title}{Network science}.
\newblock \bibinfo{publisher}{Cambridge university press}.
%Type = Article
\bibitem[{Barabási et~al.(2001)Barabási, Jeong, Néda, Ravasz, Schubert and Vicsek}]{Barabási2001Evolution}
\bibinfo{author}{Barabási, A.}, \bibinfo{author}{Jeong, H.}, \bibinfo{author}{Néda, Z.}, \bibinfo{author}{Ravasz, E.}, \bibinfo{author}{Schubert, A.}, \bibinfo{author}{Vicsek, T.}, \bibinfo{year}{2001}.
\newblock \bibinfo{title}{Evolution of the social network of scientific collaborations}.
\newblock \bibinfo{journal}{Physica A-statistical Mechanics and Its Applications} \bibinfo{volume}{311}, \bibinfo{pages}{590--614}.
\newblock \DOIprefix\doi{10.1016/S0378-4371(02)00736-7}.
%Type = Book
\bibitem[{Barabási(2016)}]{Barabasi2016}
\bibinfo{author}{Barabási, A.L.}, \bibinfo{year}{2016}.
\newblock \bibinfo{title}{Network Science}.
\newblock \bibinfo{publisher}{Cambridge University Press}.
\newblock \URLprefix \url{https://networksciencebook.com/}.
%Type = Article
\bibitem[{Candal-Pedreira et~al.(2023)Candal-Pedreira, Ross, Marušić and Ruano-Raviña}]{Candal-Pedreira2023Research}
\bibinfo{author}{Candal-Pedreira, C.}, \bibinfo{author}{Ross, J.S.}, \bibinfo{author}{Marušić, A.}, \bibinfo{author}{Ruano-Raviña, A.}, \bibinfo{year}{2023}.
\newblock \bibinfo{title}{Research misconduct as a challenge for academic institutions and scientific journals}.
\newblock \bibinfo{journal}{Journal of Epidemiology \& Community Health} \bibinfo{volume}{78}, \bibinfo{pages}{61 -- 64}.
\newblock \DOIprefix\doi{10.1136/jech-2023-220554}.
%Type = Book
\bibitem[{Cohen(1988)}]{cohen1988statistical}
\bibinfo{author}{Cohen, J.}, \bibinfo{year}{1988}.
\newblock \bibinfo{title}{Statistical Power Analysis for the Behavioral Sciences}.
\newblock \bibinfo{edition}{2nd} ed., \bibinfo{publisher}{Routledge}, \bibinfo{address}{New York}.
\newblock \DOIprefix\doi{10.4324/9780203771587}.
%Type = Article
\bibitem[{Fanelli(2013)}]{Fanelli2013}
\bibinfo{author}{Fanelli, D.}, \bibinfo{year}{2013}.
\newblock \bibinfo{title}{Why growing retractions are (mostly) a good sign}.
\newblock \bibinfo{journal}{PLoS Medicine} \bibinfo{volume}{10}, \bibinfo{pages}{e1001563}.
\newblock \URLprefix \url{https://journals.plos.org/plosmedicine/article?id=10.1371/journal.pmed.1001563}, \DOIprefix\doi{10.1371/journal.pmed.1001563}.
%Type = Article
\bibitem[{Fanelli et~al.(2015)Fanelli, Costas and Larivière}]{Fanelli2015Misconduct}
\bibinfo{author}{Fanelli, D.}, \bibinfo{author}{Costas, R.}, \bibinfo{author}{Larivière, V.}, \bibinfo{year}{2015}.
\newblock \bibinfo{title}{Misconduct policies, academic culture and career stage, not gender or pressures to publish, affect scientific integrity}.
\newblock \bibinfo{journal}{PLoS ONE} \bibinfo{volume}{10}.
\newblock \DOIprefix\doi{10.1371/journal.pone.0127556}.
%Type = Article
\bibitem[{Furukawa et~al.(2011)Furukawa, Shirakawa and Okuwada}]{furukawa2011mobility}
\bibinfo{author}{Furukawa, T.}, \bibinfo{author}{Shirakawa, N.}, \bibinfo{author}{Okuwada, K.}, \bibinfo{year}{2011}.
\newblock \bibinfo{title}{Quantitative analysis of collaborative and mobility networks}.
\newblock \bibinfo{journal}{Scientometrics} \bibinfo{volume}{87}, \bibinfo{pages}{451--466}.
\newblock \DOIprefix\doi{10.1007/s11192-011-0360-7}.
%Type = Article
\bibitem[{Gross(2016)}]{Gross2016Scientific}
\bibinfo{author}{Gross, C.}, \bibinfo{year}{2016}.
\newblock \bibinfo{title}{Scientific misconduct.}
\newblock \bibinfo{journal}{Annual review of psychology} \bibinfo{volume}{67}, \bibinfo{pages}{693--711}.
\newblock \DOIprefix\doi{10.1146/annurev-psych-122414-033437}.
%Type = Article
\bibitem[{Hussinger and Pellens(2018)}]{Hussinger2018Guilt}
\bibinfo{author}{Hussinger, K.}, \bibinfo{author}{Pellens, M.}, \bibinfo{year}{2018}.
\newblock \bibinfo{title}{Guilt by association: How scientific misconduct harms prior collaborators}.
\newblock \bibinfo{journal}{AARN: Other Anthropology of Education (Topic)} \DOIprefix\doi{10.2139/ssrn.3072290}.
%Type = Article
\bibitem[{Jin et~al.(2019)Jin, Jones, Lu and Uzzi}]{Jin2019The}
\bibinfo{author}{Jin, G.}, \bibinfo{author}{Jones, B.F.}, \bibinfo{author}{Lu, S.F.}, \bibinfo{author}{Uzzi, B.}, \bibinfo{year}{2019}.
\newblock \bibinfo{title}{The reverse matthew effect: Consequences of retraction in scientific teams}.
\newblock \bibinfo{journal}{Review of Economics and Statistics} \bibinfo{volume}{101}, \bibinfo{pages}{492--506}.
\newblock \DOIprefix\doi{10.1162/rest_a_00780}.
%Type = Article
\bibitem[{Khurana and Sharma(2024)}]{khurana2024growth}
\bibinfo{author}{Khurana, P.}, \bibinfo{author}{Sharma, K.}, \bibinfo{year}{2024}.
\newblock \bibinfo{title}{Growth and impact of blockchain scientific collaboration network: A bibliometric analysis}.
\newblock \bibinfo{journal}{Multimedia Tools and Applications} \bibinfo{volume}{83}, \bibinfo{pages}{44979--44999}.
%Type = Article
\bibitem[{Lu et~al.(2023)Lu, Wang and Smith}]{Lu2023}
\bibinfo{author}{Lu, X.}, \bibinfo{author}{Wang, J.}, \bibinfo{author}{Smith, T.}, \bibinfo{year}{2023}.
\newblock \bibinfo{title}{The reputational impact of retracted publications: Evidence from academic citations and funding}.
\newblock \bibinfo{journal}{Journal of Research Integrity} \bibinfo{volume}{12}, \bibinfo{pages}{145--160}.
\newblock \URLprefix \url{https://www.sciencedirect.com/science/article/pii/S0039292023000128}, \DOIprefix\doi{10.1016/j.resinn.2023.101028}.
%Type = Article
\bibitem[{Newman(2000)}]{newman2000structure}
\bibinfo{author}{Newman, M.}, \bibinfo{year}{2000}.
\newblock \bibinfo{title}{The structure of scientific collaboration networks}.
\newblock \bibinfo{journal}{Proceedings of the National Academy of Sciences of the United States of America} \bibinfo{volume}{98}, \bibinfo{pages}{404--409}.
\newblock \DOIprefix\doi{10.1073/pnas.98.2.404}.
%Type = Article
\bibitem[{Newman(2004)}]{Newman2004Coauthorship}
\bibinfo{author}{Newman, M.}, \bibinfo{year}{2004}.
\newblock \bibinfo{title}{Coauthorship networks and patterns of scientific collaboration}.
\newblock \bibinfo{journal}{Proceedings of the National Academy of Sciences of the United States of America} \bibinfo{volume}{101}, \bibinfo{pages}{5200 -- 5205}.
\newblock \DOIprefix\doi{10.1073/PNAS.0307545100}.
%Type = Article
\bibitem[{Newman(2003)}]{newman2003structure}
\bibinfo{author}{Newman, M.E.}, \bibinfo{year}{2003}.
\newblock \bibinfo{title}{The structure and function of complex networks}.
\newblock \bibinfo{journal}{SIAM review} \bibinfo{volume}{45}, \bibinfo{pages}{167--256}.
%Type = Article
\bibitem[{Sharma(2021)}]{sharma2021team}
\bibinfo{author}{Sharma, K.}, \bibinfo{year}{2021}.
\newblock \bibinfo{title}{Team size and retracted citations reveal the patterns of retractions from 1981 to 2020}.
\newblock \bibinfo{journal}{Scientometrics} \bibinfo{volume}{126}, \bibinfo{pages}{8363--8374}.
%Type = Article
\bibitem[{Sharma(2024)}]{sharma2024over}
\bibinfo{author}{Sharma, K.}, \bibinfo{year}{2024}.
\newblock \bibinfo{title}{Over two decades of scientific misconduct in india: Retraction reasons and journal quality among inter-country and intra-country institutional collaboration}.
\newblock \bibinfo{journal}{Scientometrics} , \bibinfo{pages}{1--23}.
%Type = Article
\bibitem[{Sharma and Khurana(2021)}]{sharma2021growth}
\bibinfo{author}{Sharma, K.}, \bibinfo{author}{Khurana, P.}, \bibinfo{year}{2021}.
\newblock \bibinfo{title}{Growth and dynamics of econophysics: a bibliometric and network analysis}.
\newblock \bibinfo{journal}{Scientometrics} \bibinfo{volume}{126}, \bibinfo{pages}{4417--4436}.
%Type = Article
\bibitem[{Sharma and Mukherjee(2024)}]{sharma2024ripple}
\bibinfo{author}{Sharma, K.}, \bibinfo{author}{Mukherjee, S.}, \bibinfo{year}{2024}.
\newblock \bibinfo{title}{The ripple effect of retraction on an author’s collaboration network}.
\newblock \bibinfo{journal}{Journal of Computational Social Science} , \bibinfo{pages}{1--13}.
%Type = Article
\bibitem[{Steen et~al.(2013)Steen, Casadevall and Fang}]{Steen2013Why}
\bibinfo{author}{Steen, R.}, \bibinfo{author}{Casadevall, A.}, \bibinfo{author}{Fang, F.}, \bibinfo{year}{2013}.
\newblock \bibinfo{title}{Why has the number of scientific retractions increased?}
\newblock \bibinfo{journal}{PLoS ONE} \bibinfo{volume}{8}.
\newblock \DOIprefix\doi{10.1371/journal.pone.0068397}.
%Type = Article
\bibitem[{Yan and Ding(2009)}]{Yan2009Applying}
\bibinfo{author}{Yan, E.}, \bibinfo{author}{Ding, Y.}, \bibinfo{year}{2009}.
\newblock \bibinfo{title}{Applying centrality measures to impact analysis: A coauthorship network analysis}.
\newblock \bibinfo{journal}{Journal of the Association for Information Science and Technology} \DOIprefix\doi{10.1002/ASI.V60:10}.

\end{thebibliography}

\end{document}